# Scientific image rendering for space scenes with the *SurRender* software

R. Brochard, J. Lebreton[*], C. Robin, K. Kanani, G. Jonniaux, A. Masson, N. Despré, A. Berjaoui
Airbus Defence and Space, 31 rue des Cosmonautes, 31402 Toulouse Cedex, France surrender.software@airbus.com
[*]Corresponding Author

**Abstract**
The autonomy of spacecrafts can advantageously be enhanced by vision-based navigation (VBN) techniques. Applications range from manoeuvers around Solar System objects and landing on planetary surfaces, to in-orbit servicing or space debris removal, and even ground imaging. The development and validation of VBN algorithms for space exploration missions relies on the availability of physically accurate relevant images. Yet archival data from past missions can rarely serve this purpose and acquiring new data is often costly. Airbus has developed the image rendering software *SurRender*, which addresses the specific challenges of realistic image simulation with high level of representativeness for space scenes. In this paper we introduce the software *SurRender* and how its unique capabilities have proved successful for a variety of applications. Images are rendered by raytracing, which implements the physical principles of geometrical light propagation. Images are rendered in physical units using a macroscopic instrument model and scene objects reflectance functions. It is specially optimized for space scenes, with huge distances between objects and scenes up to Solar System size. Raytracing conveniently tackles some important effects for VBN algorithms: image quality, eclipses, secondary illumination, subpixel limb imaging, etc. From a user standpoint, a simulation is easily setup using the available interfaces (MATLAB/Simulink, Python, and more) by specifying the position of the bodies (Sun, planets, satellites, …) over time, complex 3D shapes and material surface properties, before positioning the camera. *SurRender* comes with its own modelling tool, *SuMoL*, enabling to go beyond existing models for shapes, materials and sensors (projection, temporal sampling, electronics, etc.). *SurRender* is natively designed to simulate different kinds of sensors (visible, LIDAR, …). Additional tools are available for manipulating huge datasets ("giant textures") used to store albedo maps and digital elevation models (up to 256 TB), or for procedural (fractal) texturing that generates high-quality images for a large range of observing distances (from millions of km to touchdown). We illustrate *SurRender* performances with a selection of case studies. We place particular emphasis on a Moon landing simulation we recently computed that represents 40 GB of data and a 900-km flyby. The *SurRender* software can be made available to interested readers upon request.

**Acronyms/Abbreviations**
ADS: Airbus Defence and Space
BRDF: Bidirectional Reflectance Distribution Function
DEM: Digital Elevation Model
GNC: Guidance, Navigation & Control
GPS: Global Positioning System
GSD: Ground Sample Distance
IMU: Inertial Measurement Unit
JUICE: JUpiter ICy moons Explorer
LIDAR: LIght Detection And Ranging
LSB: Least Significant Bit
PDS: Planetary Data System
PSF: Point Spread Function
RAM: Random Access Memory
R&T: Research and Technology
VBN: Vision-Based Navigation

1. **Introduction**

Solar System exploration probes as well as in-orbit robotic spacecrafts require a high level of autonomy to perform their missions. The applications range from manoeuvers and landing around Solar System objects, to in-orbit robotic servicing missions, or space debris removal. It is an asset when manoeuvers do not rely on the ground to avoid transmission delays and also because a very high reactivity maximizes the science return of the mission during high relative velocity phases. Traditionally, Guidance, Navigation & Control (GNC) filters hybridize the input from several sensors, such as IMUs, GPS (in earth orbit), radars and altimeters. Over the last decades, it has become evident that the addition of measurements from Vision-Based Navigation (VBN) systems greatly improves the performances of autonomous navigation solutions [1, 2, 3]. ADS has developed a portfolio of VBN algorithms including relative navigation, absolute navigation, model-based navigation as well as perception and reconstruction techniques. These computer vision techniques rely on the availability of abundant test images to insure their development and validation.

Archival data from past missions can be used to test algorithms but, when they are available, they are generally sensor-specific, they lack exhaustiveness, representativeness, and often ground truth references. Some experiments are carried out on ground, thanks to test benches equipped with robotic arms handling sensors and target mock-up. Their representativeness is limited in terms of range, spatial irradiance, available scenarios, *inter alia*, and these facilities have high operating costs, so that only a few test cases can be run.

To assess performances and robustness of VBN solutions, computer generated data are highly recommended as they allow testing solutions in exhaustive cases and provide ground truth comparisons. The objective of the image processing part of the VBN, derived from computer vision technics, is often to extract information from the image geometry: track features, match edges, etc. In this context, the capability of the algorithms to extract such information depends on notions such as texture, contrast, noise, which all come from a broader field called radiometry. The same challenges exist for vision sensor design.

Some simulator softwares are well-known to the public: examples include the *Celestia* and *Blender* software or rendering engines used for video game or by animation studios. Although visually impressive, these simulators lack the realism needed for advanced image processing techniques. Typically they are designed to cope with the human vision, but they do not have the photometric accuracy that actual sensors are sensitive to.

Space applications require specialized software. For instance the *PANGU* [4, 5] utility is commonly used in the European space sector. Our team at Airbus has been developing since 2011 the image rendering software *SurRender*, which addresses the specific challenges of physical image simulation (raytracing) with high level of representativeness for space scenes. Even if it is based on classical 3D rendering techniques, it adapts those to specific space properties: high variability of objects sizes, very high range between them, specific optical properties. The software recently went through a formal qualification process, it has been used in many R&D and technological development projects for ESA (JUICE, GENEVIS), CNES, for the European Community (NEOShield-2) and internal projects (e.g. SpaceTug).

Section 2 of this paper presents the methods that compose the *SurRender* software, from first principles, to computational implementation. In section 3, we introduce the user interfaces, *the* embedded modelling language *SuMoL* and sensor models. Section 4 presents a variety of tests that have been made for its formal validation. In Section 5 we demonstrate the performances of *SurRender* by focusing on the example of a Moon landing application. In section 6 we list additional examples of applications. Finally in Section 7 we discuss future ambitions and collaboration opportunities.

## 2. Methods
### 2.1 Physical principles

*SurRender* implements the physical principles of light propagation (see Fig. 1). It solves geometrical optics equations with (backward-) *raytracing techniques*. The content of a pixel is determined by casting several rays originating from this pixel, and finding which real world objects this ray encounters until it finally intersects with a light source (Sun/stars, planets or artificial light sources). Physical optics-level effects, such as diffraction of the light by the aperture of the imaging instrument, are taken into account at macroscopic level using a Point Spread Function (PSF). The light flux is stochastically sampled within the pixels (including the probability density function defined by the PSF).

The raytracer handles multiple diffusions and reflections (This recursive raytracing technique is called pathtracing). Interaction of the light with the surface of the scene objects is modelled in terms of Bidirectional Reflectance Distribution Function (BRDF). The objects themselves can take arbitrarily complex shapes. The geometry of Solar System objects are described by Digital Elevation Maps (DEM) and spheroid models which are the basic input needed to calculate light scattering. Artificial objects are described by 3D meshes.

Providing the incoming irradiance is known, the image is naturally rendered in physical units: each pixel contains an irradiance value expressed in W/m². Providing the number of rays is large enough, the realism is ensured at subpixel level. Providing the models are accurate enough, *SurRender* images are virtually undistinguishable from actual photographs.

In its standard implementation *SurRender* produces visible (scattered light) images (although it is potentially able to make thermal infrared images, see Sec. 6.1). In addition to images, useful output is generated such as depth maps (range imaging) and normal maps (slope mapping, hazard detection and avoidance). Active optical sensors such as LiDARs can also be modelled with *SurRender* (Sec. 2.4).

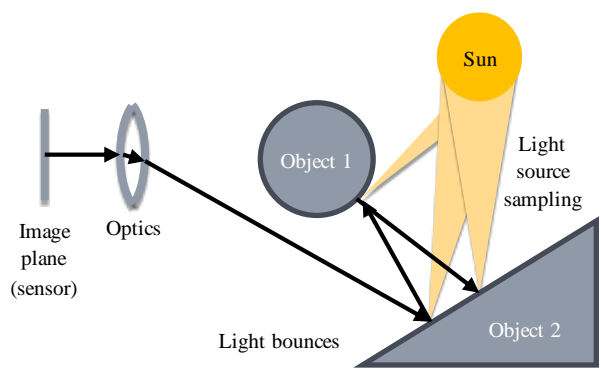

Fig. 1: *Principle of backward raytracing*

### 2.2 Computer implementation

*SurRender* enables two distinct modes of image rendering: raytracing as presented above and OpenGL (Open Graphics Library). The latter runs on GPU and allows very fast image generation, compatible with real-time processing, but at the cost of physical accuracy and fine optics/sensor modelling. This rendering mode is thus suitable for real-time setups, which do not focus on fine functional performance. In this paper we focus on the raytracing model.

The rendering engine is written in C++, it runs on CPU and on multiple cores. It is currently in its version 5. The raytracer is implemented using classical acceleration structures (Kd-tree, Bounding Volume Hierarchy) optimized for sparse scenes, as well as relief mapping with more specific ray-marching techniques [6, 7]). In addition to a number of optimizations detailed in Section 2.3, *SurRender* is characterized by a highly efficient management of cache memory.

To that purpose specific data formats were developed that are specially designed for space applications. A first pre-processing is applied to DEM to calculate *cone maps* and *height maps*. Cone maps represents, for each given point, the cone in which a ray can be traced without encountering an obstacle. Height maps represent the height with respect to a reference elevation. So-called giant textures are used to store them. Giant textures are memory-mapped, and loaded in RAM on demand: a pyramid scheme is used in such a way that only the level of details needed (given the size and resolution at stake for a given scene) is loaded. The textures can be generated from actual space mission data using common formats (PNG, JPEG, etc.) or from PDS data[*]. They can also be enriched using a *procedural fractal generator* to create additional details at small scales. This data management scheme is up to 50 times more efficient than manipulating meshes for planetary surfaces.

Computations are generally performed with double precision, which is essential for planet size objects. Indeed the elevation variations between the typical surface elevation of a Solar System object and its radius is about the order of magnitude of the LSB of floating point numbers (typically 1 m vs 1000 km). However height maps and cone maps can be stored respectively in *compact float* and *half float* formats for efficiency[†].

The size of the giant textures can far exceed the available RAM/cache memory available on the computer. To that purpose *SurRender* can be used with the FUSE software to create arbitrarily large virtual files to manage the fractal textures.

Generating an image using raytracing lasts from a few seconds to several minutes depending on the desired image quality and the scene content.

*2.3 Optimizations*

The implementation of raytracing techniques in the context of space exploration poses several challenges. *SurRender* generates images using *backward raytracing*. Raytracing is a stochastic process, in which rays are casted recursively from pixel to light source. In order to reduce the noise intrinsic to this stochastic process, a great number of rays must be traced. This process is computationally intensive, and the raytracing software must be heavily optimized to ensure the maximum image quality (lowest noise) within a minimum rendering time. Space scenes are generally sparse. Randomly sampling the volume space would therefore prove highly inefficient since most rays back-propagated from the detector pixels would not intercept any object. This difficulty is addressed by explicitly targeting the rays toward scene objects, thus avoiding a lot of unnecessary computation. The ray casting is focused in directions that contribute to the final image. Small objects are rendered efficiently by explicitly targeting them at a *sub-pixel level*. This allows simulating continuously a rendezvous mission from infinity to contact. *SurRender* also implements forward raytracing functionalities, enabling photon mapping, a process that provides maps of received surface radiances for all objects in the scene. All these optimizations – among others - enable to simulate scenes at Solar System scales.

## 3. How to use SurRender
*3.1 User interfaces*

The main application runs a *server* that can be located on the same computer as the client part, or on a remote computer. Users interact with the server using the *surrender client*. The server receives commands from the client trough a TCP/IP link, and sends the resulting image back. If sent over a network, images can be compressed to reduce the transmission duration. A specialized mode exists that emulates a camera by redirecting the data flux towards an optical stimulator for use with real sensors on test benches (*hardware-in-the-loop*) or to another client like an Ethernet camera.

For the user, *SurRender* is instantiated using external scripts. Interfaces are available in a variety of language: MATLAB, Simulink, Python 3, C++, Lua, Java. Cloud computing capabilities were recently implemented enabling a huge gain in performances through a higher level of parallelization.

The scripts describe the scene and its objects: camera, light sources, stars, planets, satellites, their positions, attitudes, shapes and surface properties. There is no intrinsic limitation in the complexity of the scene within this framework. A number of default models are provided that can easily be modified, parameterized or complemented by the user using the *SuMoL* language (see Sec. 3.2).

---

[*] "Planetary Data System", https://pds.nasa.gov.
[†] Floats are stored on 32 bits, doubles on 64 bits. Half floats and compact floats (*SurRender*-specific) are encoded only on 16 bits (reduced precision), respectively 17 bits (almost as precise as 32 bits floats).

External tools are provided in the package in order to prepare the datasets (formatting, and pre-calculation of cone maps, fractal generator, etc.). Using these tools, it is very easy to import data at usual formats such as the NASA PDS format.

*3.2 SuMoL*

*SurRender* embeds its own language, SuMoL (standing for *SurRender* Modelling Language), a high-level programming language derived from C. SuMoL is used to model various components of a simulation. The list of currently available models we provide includes:
- Projection models (pinhole, fisheye, orthographic)
- Geometrical objects (spheres, spheroids, planar & spherical DEMs)
- BRDFs (Lambertian / mate, Mirror, Hapke [8], Phong [9], Oren-Nayar [10])
- PSFs (can be field-dependent and chromatic)
- Sampling (sequential, rolling shutter, snapshot, etc.)
- Sensors properties (Section 2.5).

They implement analytical (or numerical) models, and they can be interfaced with the client "on-the-fly" i.e. without recompiling the whole software. They make *SurRender* a very versatile rendering tool, as new or specific sensor and material behaviors can be easily described by the user without changing the rendering engine. A SuMoL editor is provided as an external tool so that a user can easily implement his/her own models if needed.

*3.3 Sensor models*

By default "ideal" images are produced, not altered by any real life optical effects, *in irradiance* units (W/m²)[‡]. The images can be retrieved in greyscale or in multiple wavelengths channels: the bandwidth is specified by the user, and the BRDF and stellar spectra are chromatic. These images already account for the sensor field of view (FOV) and the size of the pixel matrix. A PSF can be specified in order to emulate diffraction by the optics.

Importantly fine sensor models can be implemented using external scripts and SuMoL wrappers. Various sensor properties and setups can be tuned: integration time, readout noise, photon noise, transmission, gain, pupil diameter, motion blur, defocus, etc. The acquisition mode can also be simulated: rolling or global shutter, push-broom, snapshot, windowing (accounting for detector motion during the acquisition). Recently the possibility of a PSF varying in the field-of-view was implemented, as well as chromatic aberrations. All these effects are natural byproducts of the raytracer. Available sensors include a "generic sensor" (classical noises and global shutter), HAS2 and HAS3 sensors [11] - which implement windowing and rolling shutter (see Fig. 2) - and more.

Active optical sensors such as LIDARs or time-of-flight cameras can also be used. The simulation includes the light emission (and reflections) from the spot or the laser.

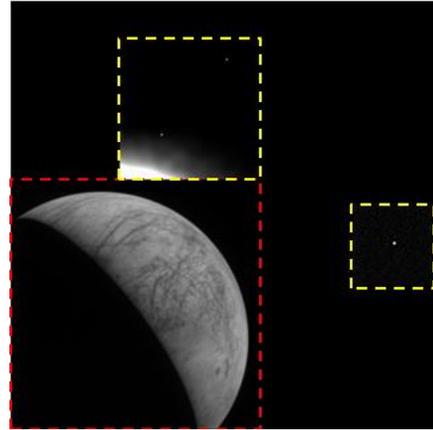

Fig. 2 : *HAS3 sensor simulation. Sensor sub-windows at different integration times enable simultaneous observation of faint stars and bright objects.*

**4. Validation and performances**

*4.1 Analytical validation*

*SurRender* has passed a formal qualification review internal to Airbus. Its physical accuracy has been tested in specific contexts that can be studied analytically, allowing quantitative assessments. Both geometric and radiometric correctness have been controlled. We verified the (subpixel) position, orientation and irradiance of objects (points, spheroids, textures) in the image plane (with pinhole model, distortions, rolling shutter), including BRDFs (Lambertian sphere or plane) and the PSF effect. The tests demonstrated exquisite accuracy and precision. The report of this analysis can be provided to an interested reader upon request.

This analytical validation was completed by a qualitative study of rendered images for complex scenes against real images.

*4.2 Validation with real scenes*

We conducted a classical test in computer graphics: the accuracy of the "Cornell box" rendering. Fig. 3 allows qualitative comparisons of *SurRender* raytracing results, with real data photographed with a CCD camera. The visual properties of the image are correctly simulated, with various optical effects such as shadows, BRDFs, inter-reflections, etc., even though in this test, pieces of information were missing (PSF, post-processing, precise camera pose, light spectrum). This particular simulation is very demanding due to the non-sparsity of the scene and the multiple reflections involved.

---

[‡] Irradiance is defined as the radiant flux received by a surface per unit area (W/m²). It corresponds to the radiance (W.m$^{-2}$.sr$^{-1}$) received from the scene, modulated by the solid angle of each pixel.

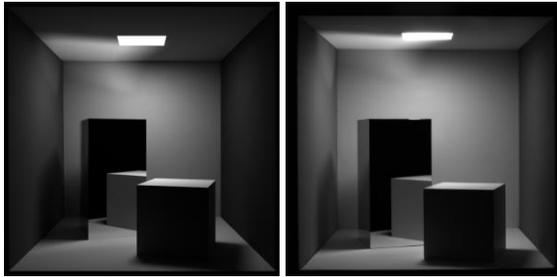

Fig. 3: *The Cornell box test. Left: SurRender image, right: real image, both at 500 nm. Note the soft shadows and the multiple light reflections between objects.*

*SurRender* has also been validated against real space images. In Fig. 4 we show a comparison with real data from New Horizon's LORRI imager. The scene includes Europa and Io. Planetshine on Io is correctly simulated. The only visible difference is the presence of volcanos around Io, which are not yet implemented. Then, in Fig. 4 we also show how the subpixel accuracy was validated against real images by showing a zoom on the limb of Ganymede, using a simple PSF model and actual data from New Horizons / LORRI.

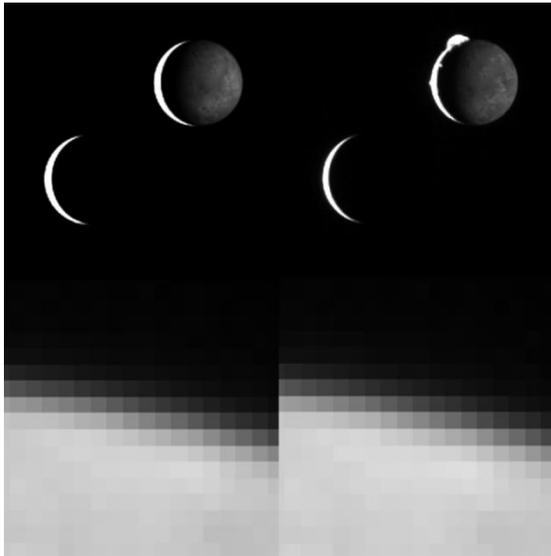

Fig. 4: *A view of the Jupiter moons from, left: SurRender, right: real images (New Horizons). Top panels: Io and Europa. Note the illumination of Io by Jupiter, and the presence of a volcano on the real image that is not yet simulated. Bottom panels: the limb of Ganymede.*

Another study was performed to qualitatively and quantitatively validate the performances against real data, this time in framework of the European Commission project NEOShield-2 [§] [12]. Simulated images were compared to actual space images from the Hayabusa / AMICA imager, and to images acquired on a test bench by industrial partners with a real camera and a mock-up of the scene. The 3D shape of the numerical model and of the mockup was constructed from AMICA images [13]. A uniform albedo and a Lambertian BRDF were used, and the illumination direction was close to the camera. A thorough analysis of the geometric precision and the radiometric precision was performed.

---

[§] *Science and Technology for Near-Earth Object Impact Prevention*. http://www.neoshield.eu/

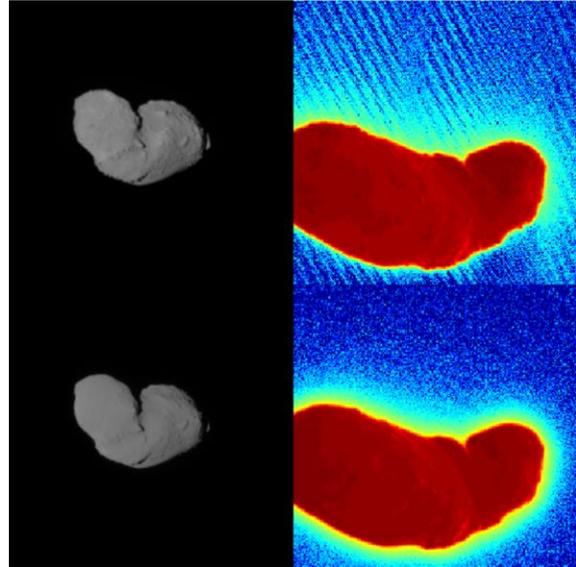

Fig. 5: *Asteroid Itokawa, as seen by the Hayabusa probe (top panels) and simulated with SurRender (bottom panels). Left panels: greyscale images; right panel: logarithmic images.*

As illustrated in Fig. 5, the visual inspection of the data is very convincing. Geometrical tests reveal error typically smaller than one pixel in average between the 3 images series based on different metrics. Radiometric tests reveal than *SurRender* and Hayabusa match within a few percent in their radiometric level. Only the residual background noise slightly differs from the actual image as can be seen on the right panels of Fig. 5. Indeed we used the AMICA official sensor specification [14], which did not document this residual pattern. The *SurRender* images are only as good as the detector model. In sum, despite model inaccuracies and realisation error, the simulated images are very close to reality. In fact, *SurRender* proved more practical and to some extent more realistic than the physical test bench as it enables to simulate a variety of scene configurations without the limitations of the laboratory. Another test consisted in observing the results of applying an IP algorithm, namely template matching, to different image sets. It was especially important as its success demonstrated the viability of *SurRender* simulations to qualify VBN algorithms.

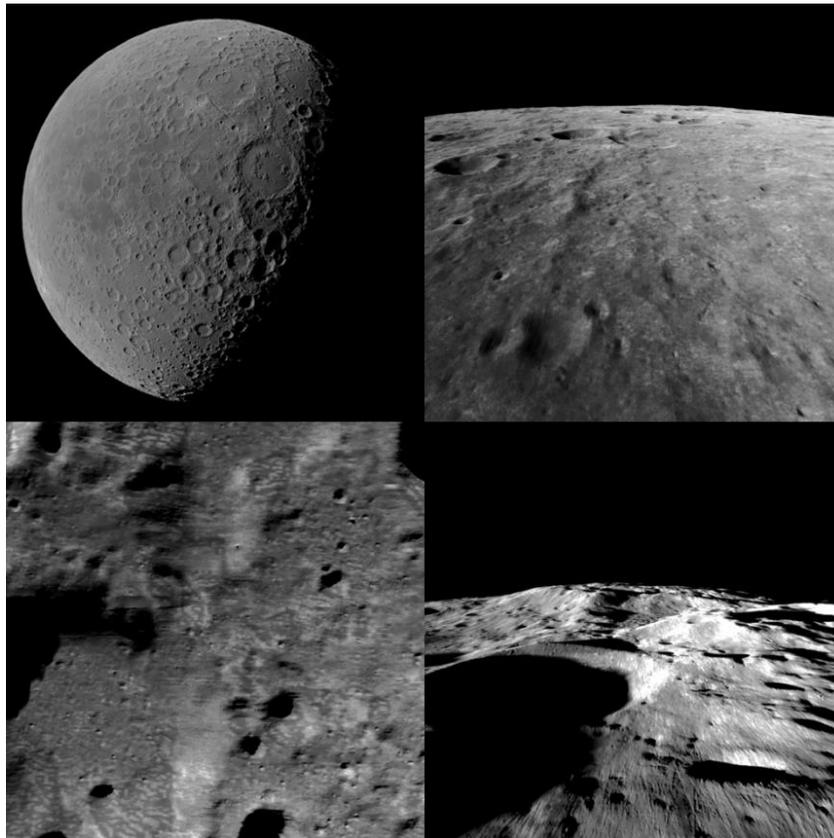

Fig. 6: *A Flyby of the Moon towards the South Pole simulated with SurRender. The rendering is made at a variety of distances: from thousands of km (top-left), a 100 km (top-right), a few tens of km (bottom-left) and finally few km (bottom-right).*

### 5. Case study: Lunar South Pole landing

Recently the software was used to tackle a difficult challenge. In the context of the ESA R&T study GENEVIS[**], simulated images were needed for a ~900 km-long flyby along a meridian of the Moon down to the South Pole. To that purpose we downloaded public data from the NASA PDS archive. We used a lunar DEM based on data from NASA's Lunar Reconnaissance Orbiter (LRO) / Lunar Orbiter Laser Altimeter (LOLA) instrument [15]. The GSD is as small as 118 m at the equator and the reported vertical accuracy is 1 m on average. In addition, we used a reflectance map at 750 nm from JAXA SELENE / Kaguya Multiband Imager [16], which has a GSD down to 237m. We assume that the BRDF of the Moon is best described by an Hapke model. The ultimate resolution that can be achieved here is limited by the data, not by the capacities of *SurRender*.

The DEM tiles are preprocessed to generate cone maps and height maps. The reflectance map has some voids that we eliminate using extrapolations with a Gaussian filter. A procedural fractal generator is used to enhance the data with additional details at smaller scales. The PDS DEM natively uses bilinear interpolations to fill gaps. Surface normals are computed using a continuous differentiation scheme in order to avoid artefacts. Intersection with the model, i.e. cone maps are computed using the step mapping technique with sub-centimeter accuracy preventing visible artefacts when viewpoint changes. After compression to half float and compact float formats, the full input weights as much as 36 gigabytes. Yet the efficient giant texture management enables to run the simulations with a framerate of about 0.2 Hz on CPU for an image size of 1024x1024 pixels and 128 rays per pixels. We used cloud computing to distribute the workload over 10 machines with 16 cores each, rendering 46.000 images in about 3 days.

For this simulation, a "generic sensor" was used. The model includes pupil and lens size, quantum efficiency, lens transmittance, gain bandwidth and integration time, readout noise, a global shutter with a certain line addressing time, etc. Photon noise and a (single) PSF are also accounted for. A selection of images is presented in Fig. 6, embracing various scales. These images have the scientific quality required for the testing of computer-vision algorithms.

---
[**] *Generic Vision-Based Technology Building Blocks for Space Applications*. Disclaimer: the view expressed herein can in no way be taken to reflect the official opinion of the European Space Agency.

## 6. Additional case studies

### 6.1 Asteroids: Itokawa

*SurRender* is particularly useful to simulate the approach of small Solar System bodies, which involves a wide distance dynamics, from far-range imaging to touchdown. We have developed a number of scenarios for asteroids based on public data, including 67P/Churyumov-Gerasimenko, Eros and Itokawa. In Fig. 7, we display images of comet 67P/CG produced with *SurRender*. At large distances, a 3D model of the asteroid is used based on a mesh shape from Hayabusa / AMICA data [13]. At close range, a DEM model of Philae landing site "Agilkia" is loaded: it corresponds to the hybridation of descent imaging and a fractal model of craters [17], and has a spatial resolution as small as 5 mm. Uniform albedo and a Hapke BRDF are assumed. For reference, the dimensions of 67P/CG's small lobe are 2.5x2.5x2.0 km, the dimensions of the large lobe are 4.1x3.2x1.3km. The sensor model consists of a simple Gaussian PSF model. The simulation ensures a continuous rendering from millions of kilometers to sub-meters distances with adequate level of details. These results were obtained in the context of the LoVa[††] project.

<hr/>

[††] *Localisation Visuelle Approche astéroïde*, ADS/CNES/LAM

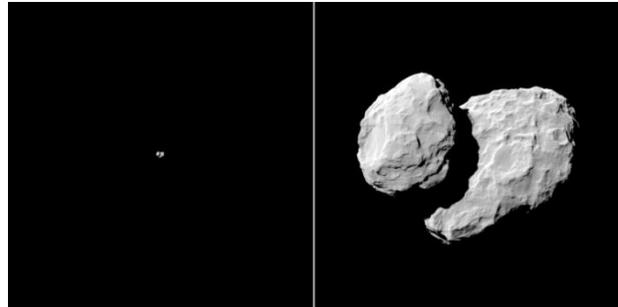

Fig. 7: *Simulated images of asteroid 67P/CP ("Chury") at thousands of kilometers then tens of kilometers.*

### 6.2 Mars (landscape)

Other useful output of *SurRender* consists of depth maps and normal maps. They are especially useful for precision landing applications, or for surface navigation (rovers) that require hazard detection functions. *SurRender* can be used to produce slope maps (or equivalently normal maps), and to simulate obstacles, or irregular terrains. In Fig. 8, we illustrate these applications on the case of Mars. This scene is generated using a digital elevation model and an albedo map assuming Lambertian BRDF. Topographic maps of Mars produced with MRO/HiRISE stereo images were used [18].

These simulations were performed for an R&T study focusing on the fusion of LiDAR and camera images. Indeed depth maps are the basis for the simulation of LiDAR acquisition. A complete simulation would also account for the laser beam propagation which is perfectly feasible. Similarly *SurRender* can be used to simulate time-of-flight cameras.

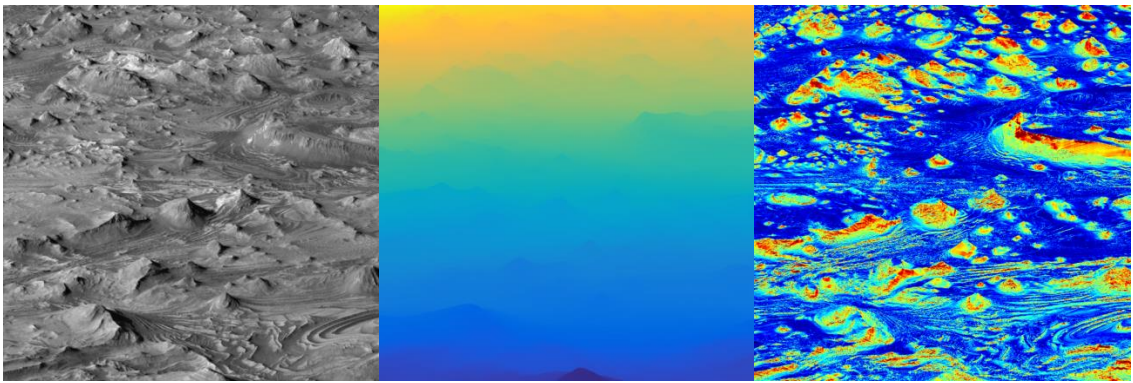

Fig. 8: *Synthetic image of the surface of Mars generated with SurRender (left) using MRO/HiRISE images and DEM and the associated depth map (center) and slope map (right).*

*6.3 The Jovian system*

*SurRender* is being used for the validation of autonomous navigation solutions for the JUICE spacecraft, ESA's future large mission that will visit the Galilean satellites of Jupiter (Europa, Ganymede and Callisto). The image processing is based on the precise localization of Jovian moons' limb during flybys [19]. To that purpose we designed high fidelity simulations of the Jovian system. The simulation includes a detailed model of the navigation camera, including multiple physical effects of the optics (blooming, chromaticity, ghosts, variable PSF) and the detector (noises, cross-talk, variable integration time). Due to the high (subpixel) precision required to render the limbs, an accurate model of the moons' BRDF is required. This problem is being tackled by Belgacem *et al*. 2018 [20] who revisit archival data from the Voyager and New Horizons spacecrafts. The results will be implemented in the ultimate simulations.

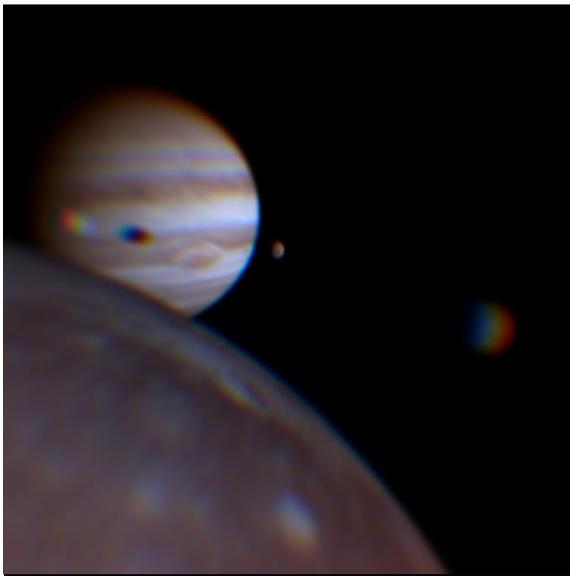

Fig. 9: *Jupiter, Europa, Ganymede and Callisto simulated with SurRender. The scene (planet positions) represents an artist's view. An actual camera model is used with exaggerated setup to highlight achromatism. The shadow of Io can be seen on Jupiter; chromatic aberrations can be seen on Callisto (right).*

Fig. 9 shows an image of Jupiter and its moons as rendered by *SurRender*. For this particular illustration, the position and size of the Moons were fixed arbitrarily based on an esthetic criterion. Jupiter and its moons are modelled with ellipsoids. The textures are composite images (2048x4096) processed from Voyager and Galileo data [21][‡‡]. The raytracer natively produces effects that were not thought ahead, for example one can notice that the shadow of Io is visible on Jupiter and has fuzzy edges where the eclipse is only partial. In a similar manner, some ghost effects may have impacted the image processing performance, but they were tackled adequately thanks to adaptations of the algorithm to these effects. Eventually these simulations will be made on a test-bench using an optical stimulator. The navigation camera will then be tested under the real sky to further validate the algorithms and the assumptions before the launch, scheduled in 2022.

*6.4 Artificial objects*

*SurRender* is also able to simulate artificial objects such as satellites and spacecrafts. It is routinely used at ADS for the design and validation of rendezvous missions including the planned in-orbit servicing vehicle *SpaceTug*, the space debris removal demonstrator *RemoveDebris* [22] and the future ESA milestone *Mars Sample Return* (MSR) [23]. It has even been used to simulate urban landscapes with a DEM, buildings and vehicles for ground-based applications. Fig. 10 shows examples of images produced with *SurRender*.

The 3D model may include moving parts, such as Solar panel, of even a robotic arm. The instrument model may also include active sensors, like in the example of *SpaceTug* depicted in Fig. 10 that includes a light spot.

The scene configuration requires a 3D model of the target: mesh models can easily be imported using standard data formats such as OBJ, 3DS, Collada formats. The main difficulty is to setup a BRDF model for each surface. Default materials available in *SurRender* such as the Lambertian BRDF (mate surface) or the Phong BRDF (plastic, MLI) properly describe satellite materials, and SuMoL enables the addition of new materials.

---

[‡‡] http://stevealbers.net/albers/sos/sos.html

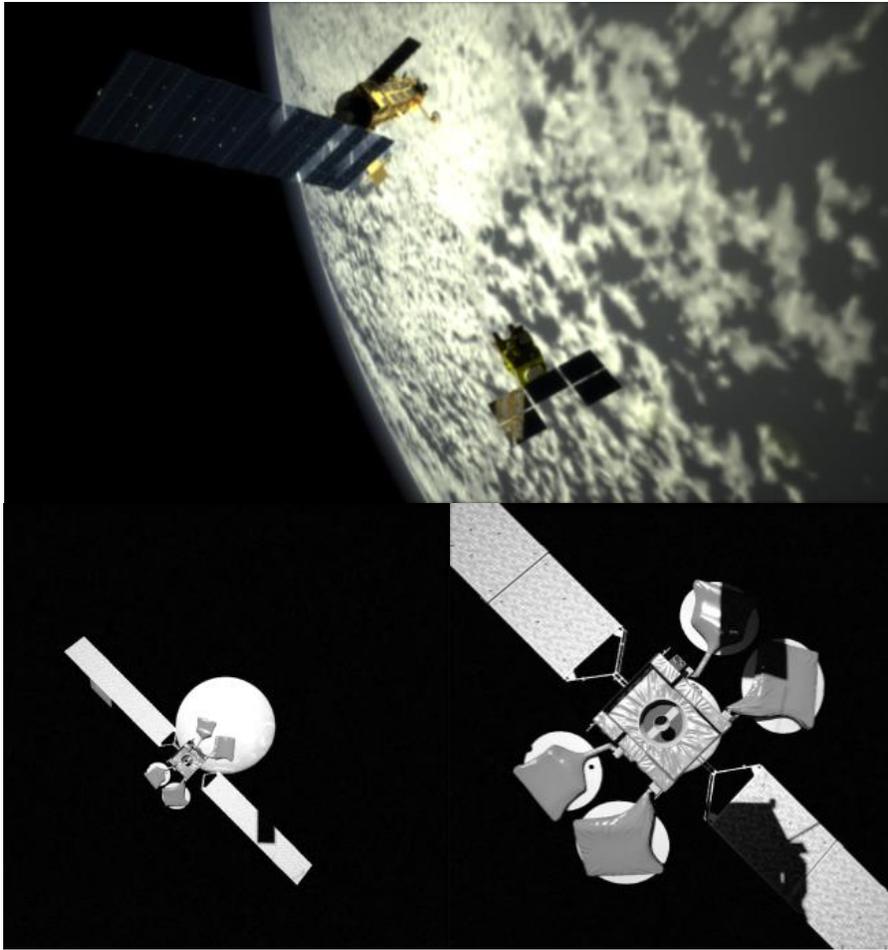

Fig. 10 *Top Panel: The SPOT 5 and ENVISAT satellites rendered with SurRender on an Earth background. Bottom panel: simulation of a rendezvous scenario. A telecom satellite is approached by the SpaceTug. Note the shadow of the tug on the satellite. On the right panel, the tug light spot enlightens the sun shadow.*

## 7. Discussion

The current capabilities of the *SurRender* software undoubtedly put it among the most powerful space simulators available on the market. Airbus and its partners have used and are using it for countless projects. Within its fields of applications, it is virtually unlimited. New scenarios and models can easily be setup and simulated in no time benefiting from the accessible interfaces and from *SuMoL*. The computer performances of *SurRender* are exquisite. It can simulate highly realistic ray-traced images in a matter of seconds (depending on scene complexity), and in realtime in *OpenGL* mode. Cloud computing capabilities are also being tested to fasten the rendering of even more complex scenes, the rendering engine will be updated to adapt to these new possibilities. *SurRender* is definitely an asset for the qualification of VBN solutions. It can also be used for sensor specification by providing representative target signal level.

### 7.1 Why SurRender?

*SurRender* is a unique tool and it offers a lot of advantages due to its very conception.

- Planetary bodies are described using geometrical model & giant textures rather than meshes. This yields a huge increase in performances in rendering time and image quality (no geometrical defects).
- Pathtracing spontaneously generates natural effects such as the secondary illumination of Io by Jupiter (Fig. 3), geometrically correct shadows / eclipses and reflexions, etc. These effects do not need to be manually setup.
- The raytracer accurately renders images with subpixel accuracy. This is due to the physical nature of the algorithm that does not rely on any spatial sampling (unlike OpenGL).
- Similarly it naturally supports temporal detector sampling (rolling shutter, pushbroom).
- *SurRender* runs on CPU allowing virtually unlimited memory use, notably using cloud computing.

- *SurRender* is designed to simulate a continuous approach with distances varying by several orders of magnitude, with a constant SNR and without any threshold effects.
- The PSF implementation obeys the laws of physical optics and naturally accounts for the PSF trail to simulate effects such as blooming even from out-of-field objects.
- SuMoL is a versatile add-on that gives limitless possibilities to the user in terms of scene description, sensor design, etc.
- Real-time rendering is supported through OpenGL with the same scene description/models as the raytracer.

*7.2 Future improvements*

At the moment *SurRender 5* is designed to render optical (visual/near-infrared) images, it does not natively handle thermal infrared images. Yet it is possible to input emission efficiencies to mimic infrared light. At this stage only equilibrium temperature can be modelled that way (no time dependence). It is perfectly possible to couple the software with a thermic simulator that would provide temperature or emission maps. In future versions, these features could become standard.

Furthermore we are planning to develop LIDAR modelling capabilities. Currently *SurRender* produces essentially depth maps, and it can include illumination for the spot or laser. Some work is needed (*SuMoL* interfaces, optical paths) to increase the realism, for example by simulating complex equations of light propagation to generate interferences (speckles). We are also testing the ability of *SurRender* to handle moving parts, such as a robotic arm. Although feasible, this poses some challenges (computing performance, conventions for cinematics) that are the topic of new R&T projects.

The current limitation for Earth applications is the lack of a model for transparent media. Therefore atmospheres cannot be accurately rendered. Yet they can be mimicked using the BRDF of an hologram, an approximation that holds as long as the observer is located outside of the atmosphere. This addition will make it able to simulate (semi-) transparent media like lenses, oceans, ice (subsurface diffusion), etc.

*SurRender* does not simulate the path of light through a refractive or reflective optics. Instead, a projection model is used and some limitations exist owing to the optical models. We are continuously improving *SurRender* to include more physical effects.

*7.3 Collaboration opportunities*

The remaining limitation of *SurRender* is the available manpower to import new data and develop always more scenarios. Our teams and partners develop new scenes and models whenever motivated by a project. For Solar System exploration, we have developed models and scenes for the Moon, Mars, the Jovian system, Itokawa, Eros, 67P/CG. *SurRender* is being used for in-orbit applications such as *SpaceTug* and *RemoveDebris,* for *MSR* and for *ExoMars*. Various sensor models are already available and we are building a camera model that will include a great number of optical defects, as well as active sensors. The limit is the detail level of the detector and object models themselves. We are open to collaborations to enrich the possibilities of *SurRender*. An instrument manufacturer or an academic could use SurRender to test his instrument or scene model, for example to generate new scenes or even write a complementary scene generator. The interfaces are easily accessible to a trained engineer or scientist and it is compatible with standard data formats. Interested readers are invited to contact us via the *SurRender* website: *www.airbus.com/SurRenderSoftware.html*.

**Acknowledgements**

We thank the institutions and partners who placed trust in *SurRender* enabling to support its development, in particular the European Space Agency (ESA), the Centre National d'Etudes Spatiales (CNES) and the European Commission (EU). Some results presented in this paper were carried out under a programme of and funded by the European Space Agency (ESA Contract No. 4000115365/15/NL/GLC/fk). Some results were carried out under CNES R&T program R-S16/BS-0005-032 and some other under EU's Horizon 2020 grant agreement No 640351.

This paper was presented for the 69th International Astronautical Congress (IAC), Bremen, Germany, 1-5 October 2018. Contribution reference: IAC-18,A3,2A,x43828. Published by the IAF, with permission and released to the IAF to publish in all forms.